小説の可視化について
―与謝野晶子の『源氏物語』の訳本を例として

About how to make novel Visible
by using *Newly Translated Tale of Genji*
as an example

# 要　旨

　本文は、源氏物語の可視化を目的として、自然言語処理技術を使用し、統計解析・感情分析等の方法により、原文に対応し、情報学の角度から再び小説の内容を発掘した。

　小説の可視化の基本方法をまとめにし、研究手段およびその内容は以下の通りである。

　第一章では、頻度統計に基づき、データマイニングによって TF-IDF 運算を導入し、与謝野晶子の『源氏物語』訳本のキーボードを発掘した。言わば、最も重要な単語を発掘することである。

　第二章では、単語の感情極性から、全文の感情を分析することである。更に、作品の文風並びにその美学的な価値を考察し、全文に溢れた感傷的な基調に対応した。

　第三章では、単語のネットワークおよび多元化尺度構成法を使用して、文章の内容を分析することである。特に単語の関係とその使用状況を中心として研究した。それに、宇治十帖の著者に関わることを探究した。

**キーワード**　小説の可視化　自然言語処理　統計解析

# Abstract

This paper aims to make *Tales of Genji* visible by using natural language processing, mathematic analysis, emiton analysis. Based on novel, mining data from content of this novel at respect of information abstracting.

Summing up the fundamental method of novel visualization, our work are as follows:

- Based on frequency analysis, we use tf-did to abstract keyword of *Newly Translated Tale of Genji*, which means the most important word in each chapter.

- We recognize the emotion of word to analysis the emotion of each chapter of *Newly Translated Tale of Genji*. Next, we think about the connection between the result of emotion analysis and literature analysis, showing we can get same result by natural language processing.

- We build a network of all the word apperanced in *Newly Translated Tale of Genji*. Make a study of relationships between words. Further, we search the writer of *Uji Chapters*.

# 目　次

# はじめに

　文学作品に対する評価は元々十人十色だと言える。例えば、評者が自信満々で分析結果を発表する時には、「客観的なものではない」という質疑が常にある。最近、情報学分野で新たな方法の応用が広がり、その中、斬新な研究手段とした可視化手法を使用し、文学の分野で古典作品を解析することが可能になった。例えば、井波らによるウィーブレット解析を使った『源氏物語』や『銀河鉄道の夜』における感情の変化の可視化[1]、山田・村井らによるシェイクスピア作品のストーリーの可視化[2]、細井による『源氏物語』の統計解析のためのテキストマイニング[3]が行われた。彼らの研究により、小説の可視化方法の道を明けさせたが、文学作品となると、どのような具体的な解析方法を運用することが研究者に困らせる。特に、古典的な作品の内容からみると、感情の起伏および本筋と共に発展される特徴がある。ゆえに本稿の研究は、感情の起伏と物語の仕組みを重要なテーマとして進められたのである。

　『源氏物語』は、平安時代中期貴族の生活を描写する作品である。主人公源氏の一生を通じて、その恋愛、栄光、没落、政権への野心と権力闘争を基にして、平安時代の貴族社会が描かれた。『源氏物語』の原文は、54帖からなる長編小説である。およそ100万文字、２２万文節で、500名余りの人物が登場し、70年近くの出来事が描かれた。また、800首くらいの和歌を含んで、典型的で広大な王朝物語である。全書の構成となると、各帖から三部に分けることができる。文献初出が1008年であったが、成立時間はまた不明である。「世界最古の長編小説」と言われても、その地位が疑われるが、「古典の中の古典」の評価に恥じない。そのため、国文学に関する多くの研究は名前負けであったが、『源氏物語』について研究の価値がもちろんである。

　本稿では、統計解析や自然言語処理の方法を使用し、小説内容を通じ、その

---

原文を可視化手法で処理した。具体的にいえば、第一章では、伝統的な手法を行い、文章の基本的な特徴を発掘した。第二章では、感情分析手法を運用し、文字から作者の感情を復元した。第三章では、文章の文脈と仕組みを実際に可視化にした。

# 第一章 伝統的な統計方法の解析方法

第一節 概要

　文学と言うのは、作家の構想によって、世界のイメージを具体的に描画するものである。文学評論家や読者は自分の暗黙知を参照し、自らの言葉遣いを作り出す。その過程を抽象化にして、その本質は文学作品の文体、文法構成、作品の暗黙知などの文字特徴から、自分の思考まで巻き込んで作品を理解することである。本章の目的は、作品に書かれた語、文、文章に基づき、その中に潜んだ共通点を探究して分析することである。最後にグラフで研究結果を表す。

第二節 頻数分析

　統計は、ある現象の調査結果によりデータを把握することである。また、本文でのデータというのは、調査を基にして獲得する数量データである。単語の頻度があながち単語の重要性を如実に反映し限らず、だが、文章のキーワードを指示して、一定の役となると思われる。本章では、Mecabを使い、単語間の区切りおよび形態素を解析した。その後、キーワード抽出の運算で単語の重要性を評価した。言わば、頻度分析とTF-IDFの運算方法を使用し、キーボードを発掘して分析することである。

第三節 Mecab工具

　Mecabはオープンソースの形態素解析エンジンで、奈良先端科学技術大学院大学出身の工藤拓によって開発された。工藤拓はGoogleソフトウェアエンジニアでGoogle日本語入力開発者の一人である。このエンジンの名称が開発者の好物「和布蕪（めかぶ）」から取られた。パラメータの推定については、Conditional Random Fields (CRF)が用いられる。

第四節 TF-IDF 方法

TF-IDFは、文書中に含まれる単語の重要度を評価する手法の1つであり、主に情報検索やトピック分析などの分野で用いられている。TF-IDFは、TF（Term Frequency、単語の出現頻度）とIDF（Inverse Document Frequency、逆文書頻度）の二つの指標に基づいて計算される。

$$tfidf_{i,j} = tf_{i,j} \cdot idf_i \qquad (1)$$

$$tf_{i,j} = \frac{n_{i,j}}{\sum_k n_{k,j}} \qquad (2)$$

$$idf_i = \log \frac{|D|}{|\{d : d \ni t_i\}|} \qquad (3)$$

$N_{i,j}$は文書$d_j$における単語 ti の出現回数、$\sum_k n_{k,j}$は文書$d_j$における全部単語の出現回数の和、$|D|$は総文書数、$|\{d:d \ni t_i\}|$は単語$t_i$に含んだ文書数である。単語の重要度を数値化に転換することは重要だが、特に、文の構成要素となって、頻繁に出現する単語（例えば、格助詞や助詞）の重要度は適当的に下げねばならない。

第五節 原文のダウンロード

インターネットの発展によって、便利に数多くの資料を探し出すことが可能になった。また、文学名著の資料を捜索するたび、書面で作成されたファイルが唯一ではないと思われる。古文原文の研究については、米国議会図書館アジア部日本課が所蔵する『源氏物語』は、塗り箱入りの非常に美しい姿で伝わってきた古写本である。写本が室町時代から江戸時代に製造された。本論文の著者がpythonを使って、文章の内容を解析し、utf-8・txtの格式でセーブした。この中、タイトルという原文内容に関係ない部分を削除した。原文の句読点にかかわらず、原著が総計18764文からなるもので、まとめて1371kbである。『源氏物語』の原文がデートマイニングの対象として、古文で作成されたので、

古典文法処理の難度がより高くて、本文では与謝野晶子の現代語訳を使用した。

第六節 単語間の区切り

キーワードの検索を行い、本論文の著者は予め作品の全単語を指令で全自動に配列に変換して記録した。日本語では単語間にはっきりした区切りがないので文字が連続する。単語の分解において、高度な自然言語処理技術が必要だ[4]。例えば、「『源氏物語』を可視化しましょう」という文を分解し、「『　源氏　物語　』　を　可視化　しま　しょう」という形式に変化する。その形式はコンピューターで分析することができる。技術の導入により、意味を理解せずに単語を分解することが可能だと思っている。故に、本研究では Mecab と Python を使って、『源氏物語』を分解する。

『源氏物語』の現代語訳を底本として、形態素解析が行われる。形態素解析とは、文法的な情報の注記の無い自然言語のテキストデータから、対象言語の文法や、辞書と呼ばれる単語の品詞等の情報に基づき、形態素の配列から分割し、それぞれの形態素の品詞等を判別する作業である。

その結果は以下の通りだ。

---

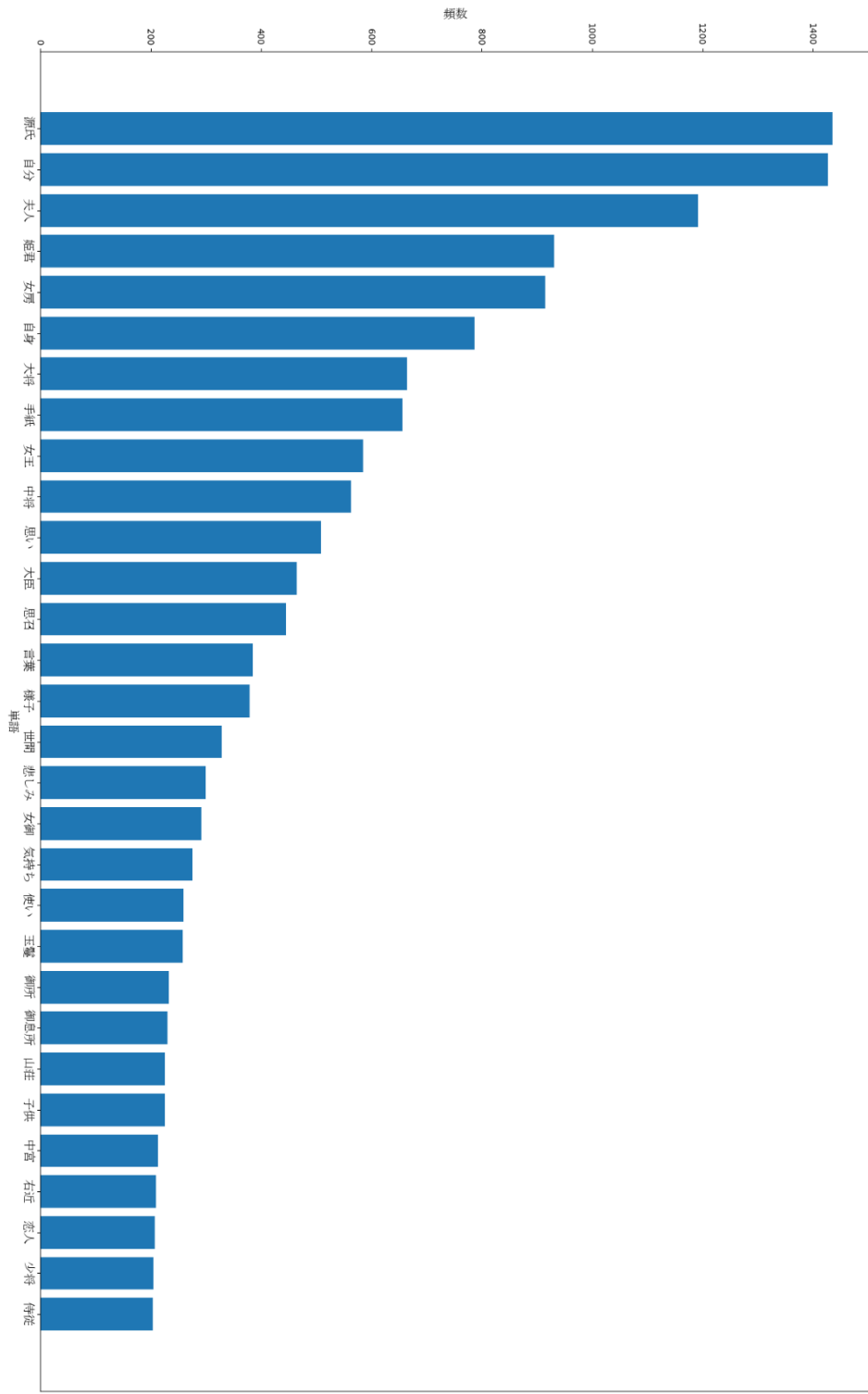

名詞の頻度（上位の30）

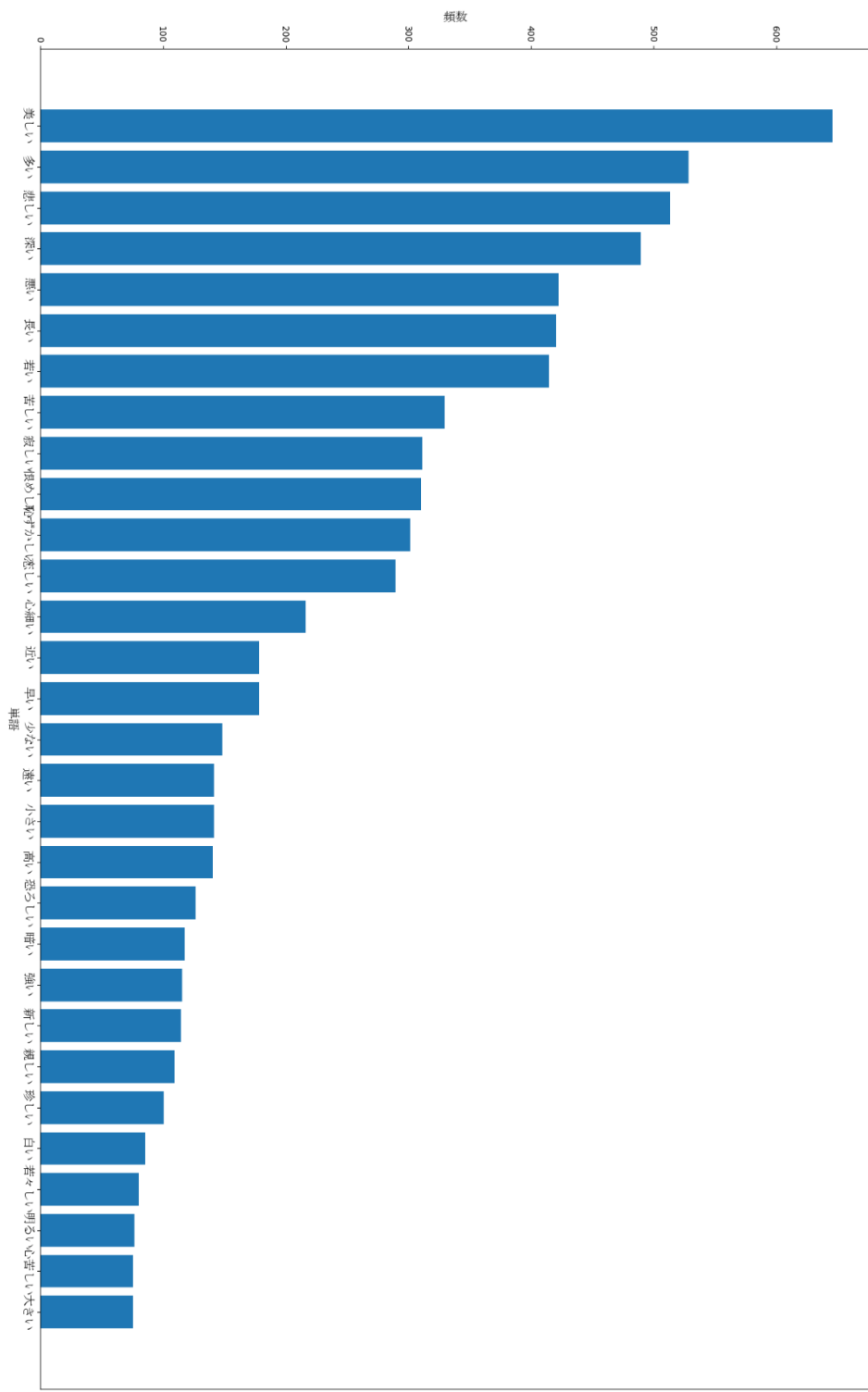

形容詞の頻度（上位の30）

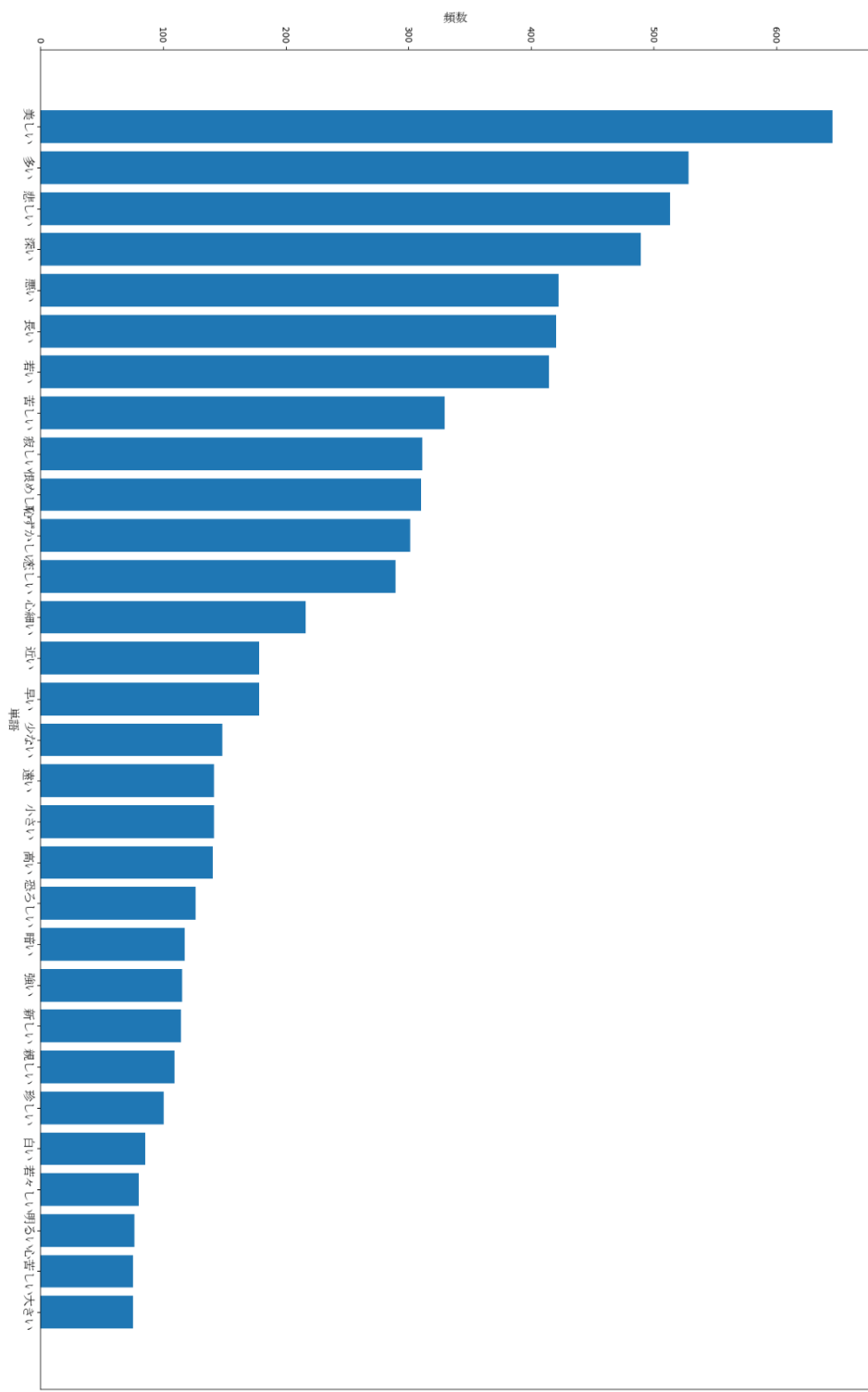

形容詞の頻度（上位の30）

# 第二章 感情分析

　心を動かされる文字を読むたびに、文字に潜んだ作家の感情と共鳴して、胸の奥に響かせる。言い換えれば、感情移入と言われる。感情表現の研究においては、その表現を抽象する方法が必要である。その技術を使用し、文章や文体の表現形式に注目して、全文の「キーワード」を選択される。解析者の主観的な選択ではなく、コンピューターの処理により、客観的な結果が得られる。その点からみると、従来指摘される文体論の問題にかかわる客観性と全体性の提示も可能になった。また、選択されたキーワードを解析し、作品の暗黙知も可視化になれる。

## 第一節 単語の感情極性

　単語の感情極性とは、ある単語が良い印象を持つかどうか、その語感を指すことである。判定基準は二値変数で作られた単語感情極性対応表である。具体的に言えば、例えば、「美しい」はポジティブであり、「悲しい」はネガティブである。単語の感情極性を判定する方法の研究は、文書内に含まれる感情を発見することをはじめ、様々な応用問題にかかわる重要な研究である。今回は東工大の奥村・高村研究室感情極性を電子のスピンの方向と見なし、シソーラス、コーパスによって構成されたネットワークをスピン系でモデル化にした。論文によって、３０００語あまりの単語を種とした場合は約９０％正解率単語の感情極性判定を実現した。

　感情分析のプログラムは Mecab の単語区切りの結果に基づいた。感情分析に使う辞書と言えば、本論文の著者が今回使ったのは、東工大の高村教授が作って公開された「PN Table」[5]という辞書である。この辞書で、単語に対応する極性情報が−1〜+1 の間で割り当てられており、−1 に近いほどネガティブ、+1 に近いほどポジティブだと判定される。後述のように、ゼロがニュートラルと判定されたのは良いのかどうか、それについては、更に討究が必要だと思って

---

[1]高村大也，乾孝司，奥村学．スピンモデルによる単語の感情極性抽出［J］．情報処理学会論文誌，2006，47(2)：627-637.

いる。本文では、ゼロがニュートラルという観点を採用した。

本文では『源氏物語』の現代語訳を句点で区分し、それぞれの文のP/N数値計算を行った。「PN Table」を使い、全てのP/N数値を統計してから0.025の間で区分した。以下は頻度分布図である。

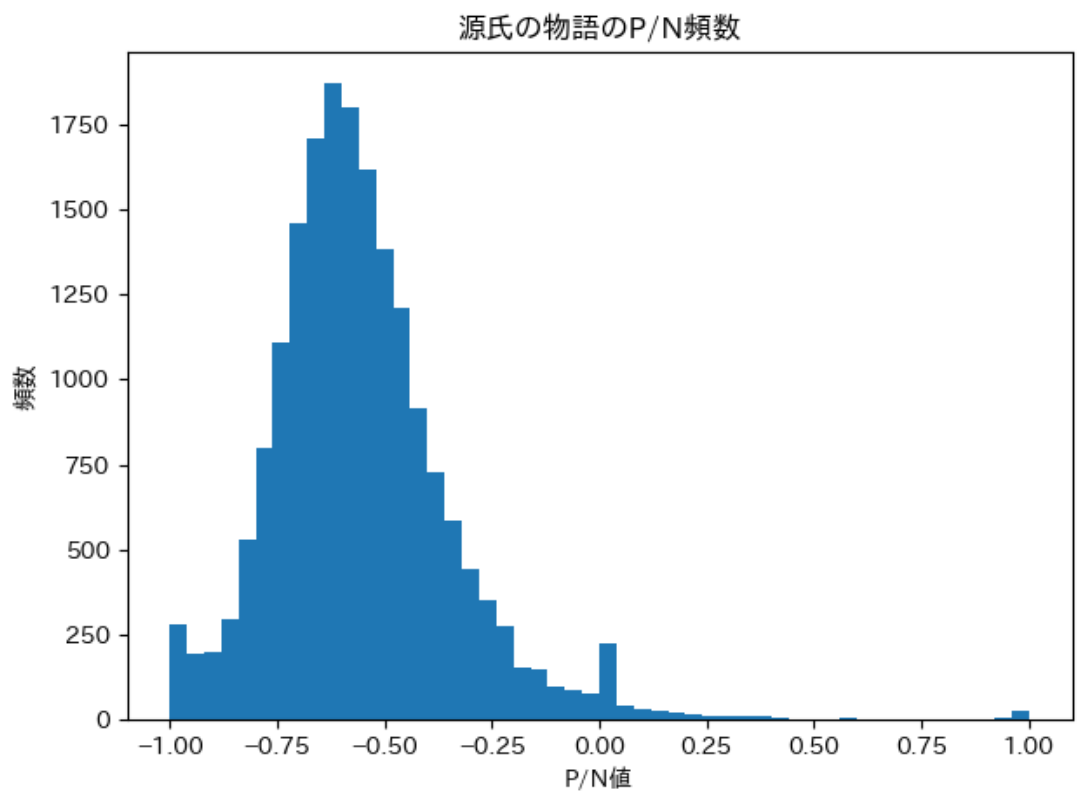

結果により、『源氏物語』全編の単語がネガティブという傾向がある。具体的には、約９５％以上の内容がネガティブだと見える。「全文を分析し、感情の起伏を探究する」を目的として、本研究は各章のP/N値を抜け出すことである。

結果によると、作品の文頭では、感情がネガティブで、天皇が藤壺氏に寄せて切なく思念を描写された。この悲恋に伴いのは、哀傷な雰囲気だと推察される。だが、源氏の青少年時代を描いた部分、文章の内容はポジティブに変化した。二十章のあたり、すなわち「明石の章」では、源氏が紫姫と分離し、文章の感情が再びネガティブに陥った。終わりまで、哀愁という感情が続いていた。この感情起伏がストーリーの脈絡に応じたと思われる。

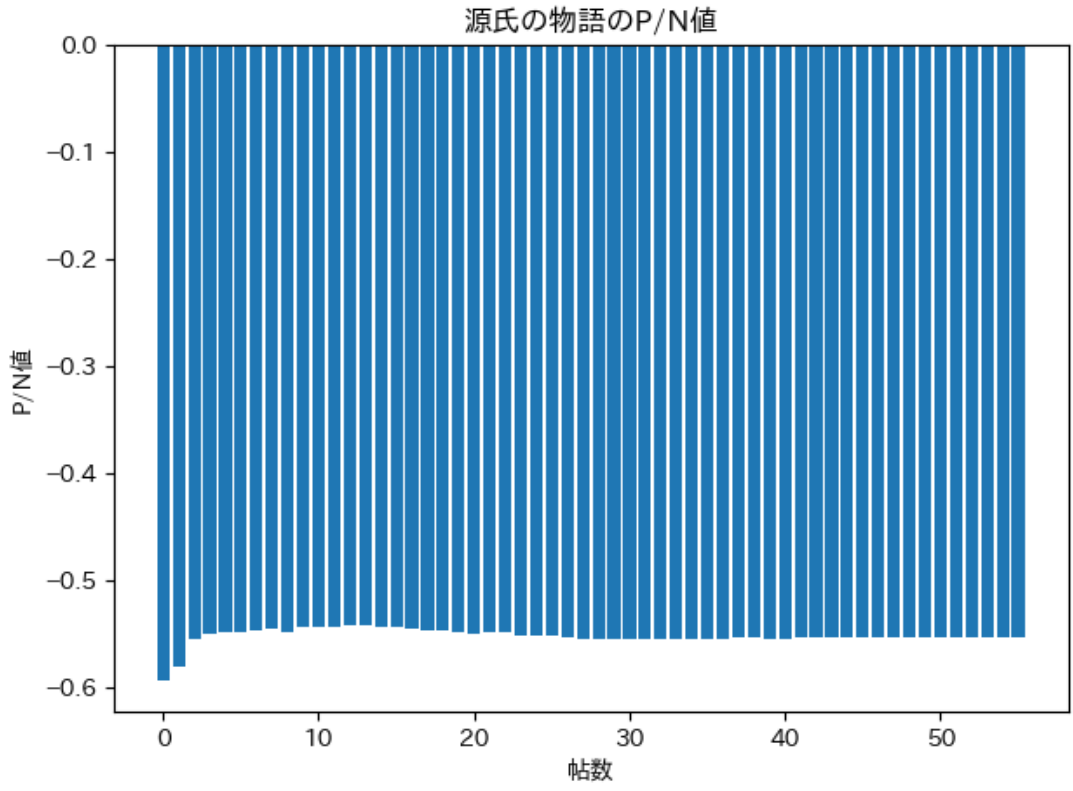

源氏の物語のP/N値

# 第三章 文章可視化

## 第二節 単語のネットワーク

これは『源氏物語』の現代語訳に基づいて生成した単語のネットワークである。単語の位置や性質によって、単語のネットワークからCommunityが作成されていた。単語それぞれの対応するサークルがあり、サークルの大きさが言語の重要性を表明されて、サークルの距離が関連性を表示される。異なるコミュニティの単語が異なる色をづけられた。例えば、「源氏」が黄色の02コミュニティの中、「中将」と近づくと、その関係の近さを表明する。あるいは、緑色の01コミュニティでは、「薫」と「姫君」があり、宇治十帖の主人公源氏でも包含された。帝と院共に源氏の父親に対応し、光源氏は悲劇のヒロインだと推察される。また、この悲恋に繋がる手紙が独立したコミュニティとなった。この点からみれば、当時社会の風貌が見えると思われる。

## 第三節 多次元尺度構成法

　多次元尺度構成法は多変量解析の一手法である。主成分分析の様に分類対象物の関係を低次元空間における点の布置で表現する手法である（相似する言葉は近くに設置され、相似しない言葉は遠くに配置される）。

　結果によると、宇治十帖と前文との関係がある。また、言語間の関係を探求すれば、言語の共通点がないと考えられる。多次元尺度構成法によっては、必ず精確な数値を提供するかぎらない。だが、言語の重要性に関しては、ある程度の証明力があると思われる。

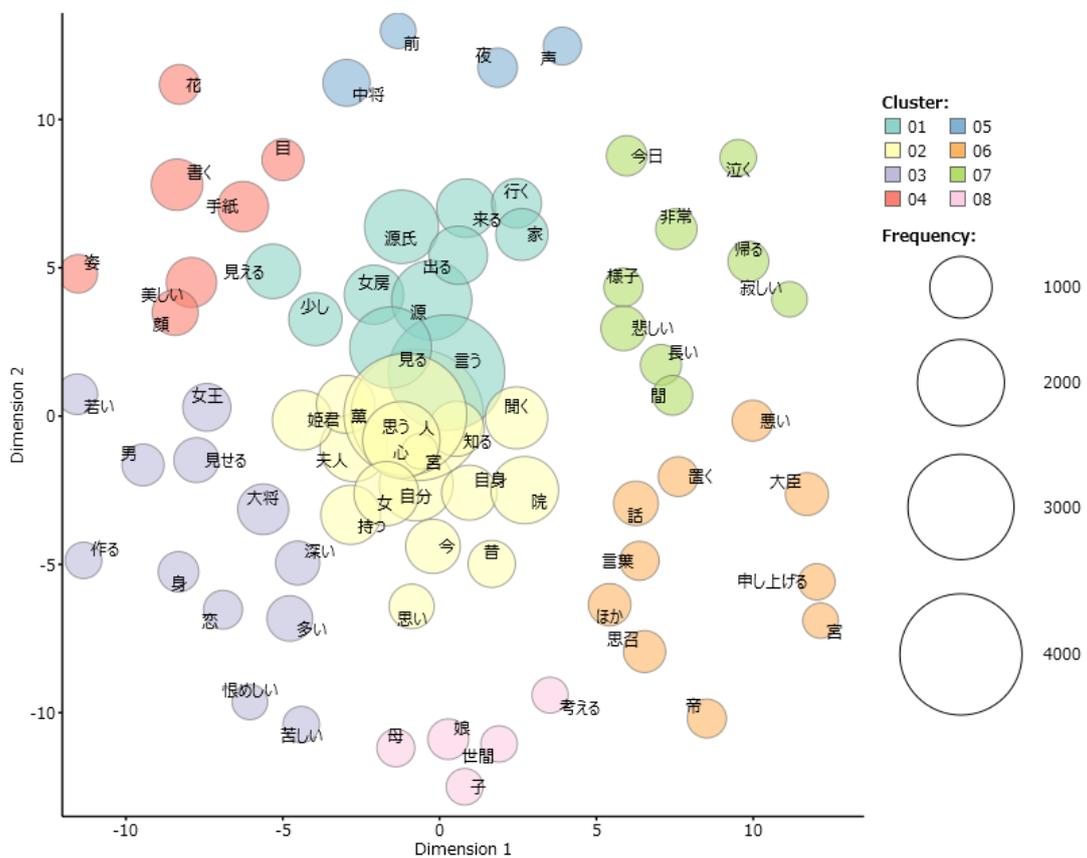

# 終わりに

　本文は、主に『源氏物語』の現代語訳を例として、可視化の技術を利用し、この小説を解読した。統計分析・感情分析・文章可視化という三つの点から全文を研究し、新たな視点で文章を解読した。

　より体系的なデータを獲得して、主にはグラフで研究結果を展示した。

　可視化の分析手段に関しては、現有の Mecab と「PN Table」辞典を参考として使用した。

# 参考文献

附录

表 源氏物语中频数排名前 100 的名词、动词和形容词

| 順位 | 名詞 | 頻数 | 動詞 | 頻数 | 形容詞 | 頻数 |
|---|---|---|---|---|---|---|
| 1 | 源氏 | 1434 | 思う | 4713 | 美しい | 645 |
| 2 | 自分 | 1426 | 言う | 3630 | 多い | 528 |
| 3 | 夫人 | 1191 | 見る | 1759 | 悲しい | 513 |
| 4 | 姫君 | 930 | 聞く | 1007 | 深い | 489 |
| 5 | 女房 | 914 | 持つ | 919 | 悪い | 422 |
| 6 | 自身 | 786 | 来る | 894 | 長い | 420 |
| 7 | 大将 | 663 | 出る | 885 | 若い | 414 |
| 8 | 手紙 | 655 | 知る | 784 | 苦しい | 329 |
| 9 | 女王 | 584 | 見える | 774 | 寂しい | 311 |
| 10 | 中将 | 562 | 書く | 681 | 恨めしい | 310 |
| 11 | 思い | 507 | 行く | 631 | 恥ずかしい | 301 |
| 12 | 大臣 | 464 | 見せる | 511 | 恋しい | 289 |
| 13 | 思召 | 444 | 帰る | 414 | 心細い | 216 |
| 14 | 言葉 | 384 | 置く | 398 | 近い | 178 |
| 15 | 様子 | 378 | 作る | 330 | 早い | 178 |
| 16 | 世間 | 328 | 申し上げる | 330 | 少ない | 148 |
| 17 | 悲しみ | 299 | 泣く | 329 | 遠い | 141 |
| 18 | 女御 | 291 | 考える | 322 | 小さい | 141 |
| 19 | 気持ち | 274 | 出す | 305 | 高い | 140 |
| 20 | 使い | 258 | 死ぬ | 300 | 恐ろしい | 126 |
| 21 | 玉鬘 | 257 | 逢う | 296 | 暗い | 117 |
| 22 | 御所 | 232 | 愛す | 289 | 強い | 115 |
| 23 | 御息所 | 229 | 立つ | 275 | 新しい | 114 |
| 24 | 山荘 | 225 | 申す | 261 | 親しい | 109 |
| 25 | 子供 | 225 | 得る | 256 | 珍しい | 100 |
| 26 | 中宮 | 212 | 覚える | 232 | 白い | 85 |
| 27 | 右近 | 209 | 変わる | 224 | 若々しい | 80 |
| 28 | 恋人 | 206 | 思い出す | 223 | 明るい | 76 |
| 29 | 少将 | 204 | 受ける | 222 | 心苦しい | 75 |
| 30 | 侍従 | 203 | 生きる | 216 | 大きい | 75 |
| 31 | 座敷 | 197 | 寝る | 207 | 古い | 68 |
| 32 | 相手 | 189 | 忘れる | 206 | 見苦しい | 66 |
| 33 | 乳母 | 189 | 感じる | 202 | 情けない | 65 |
| 34 | 尚侍 | 187 | 悲しむ | 195 | 惜しい | 64 |
| 35 | 物思い | 184 | 送る | 191 | 重い | 63 |
| 36 | 態度 | 179 | 話す | 189 | 短い | 63 |
| 37 | 中納言 | 176 | 引く | 179 | 濃い | 61 |
| 38 | 僧都 | 175 | 起こる | 179 | 低い | 60 |

| 39 | 人々 | 172 | 捨てる | 169 | 怪しい | 54 |
| --- | --- | --- | --- | --- | --- | --- |
| 40 | 音楽 | 169 | 取る | 167 | 頼もしい | 48 |
| 41 | 感じ | 169 | 弾く | 167 | 貴い | 47 |
| 42 | 良人 | 169 | 困る | 163 | 弱い | 46 |
| 43 | 東宮 | 166 | 暮らす | 161 | 優しい | 44 |
| 44 | 若君 | 160 | 離れる | 161 | 堪えがたい | 43 |
| 45 | 一つ | 159 | 隠す | 160 | 気高い | 43 |
| 46 | 御覧 | 159 | 訪ねる | 160 | 尊い | 43 |
| 47 | 好意 | 159 | 続く | 154 | 軽い | 42 |
| 48 | 御簾 | 157 | 似る | 151 | 醜い | 42 |
| 49 | 人生 | 157 | 立てる | 151 | 淡い | 42 |
| 50 | 浮舟 | 152 | 住む | 149 | 赤い | 41 |
| 51 | この世 | 151 | 寄る | 148 | 女らしい | 39 |
| 52 | 時代 | 148 | 語る | 148 | 柔らかい | 38 |
| 53 | 宮中 | 140 | 残る | 148 | 広い | 37 |
| 54 | 機会 | 139 | 思ふ | 146 | 快い | 35 |
| 55 | 居間 | 138 | 入れる | 146 | 済まない | 34 |
| 56 | 几帳 | 138 | 聞こえる | 140 | 浅い | 34 |
| 57 | 運命 | 135 | 終わる | 138 | 荒い | 32 |
| 58 | 故人 | 135 | 恨む | 136 | 重々しい | 32 |
| 59 | 人間 | 134 | 着る | 136 | 清い | 32 |
| 60 | 女性 | 133 | 笑う | 135 | 薄い | 31 |
| 61 | 惟光 | 129 | 読む | 134 | 痛い | 29 |
| 62 | 考え | 129 | 吹く | 131 | 久しい | 28 |
| 63 | 性質 | 129 | 呼ぶ | 130 | 軽々しい | 28 |
| 64 | 帰り | 125 | 待つ | 129 | 堅い | 28 |
| 65 | 左大臣 | 122 | 惹く | 128 | 細長い | 27 |
| 66 | 大納言 | 119 | 違う | 127 | 悩ましい | 27 |
| 67 | 内大臣 | 117 | 与える | 127 | 忙しい | 27 |
| 68 | 奥様 | 115 | 迎える | 124 | 涼しい | 27 |
| 69 | 夫婦 | 115 | 慰める | 122 | 細い | 26 |
| 70 | 御殿 | 113 | 歌う | 121 | 弱々しい | 26 |
| 71 | 宰相 | 111 | 散る | 119 | 寒い | 25 |
| 72 | 主人 | 109 | 行なう | 117 | 狭い | 25 |
| 73 | 身分 | 109 | 認める | 114 | 賢い | 25 |
| 74 | 秘密 | 109 | 生まれる | 111 | 正しい | 25 |
| 75 | 感情 | 108 | 伺う | 110 | 憎い | 24 |
| 76 | 気分 | 104 | 許す | 107 | 幼い | 24 |
| 77 | 世の中 | 104 | 亡くなる | 107 | 厚い | 23 |
| 78 | 仰せ | 102 | 仰せる | 106 | 騒がしい | 23 |
| 79 | 勤め | 102 | 選ぶ | 104 | 睦まじい | 22 |
| 80 | 恨み | 102 | 別れる | 104 | 黒い | 21 |
| 81 | 住居 | 102 | 頼む | 103 | 思いがけない | 21 |

| 82 | 場所 | 102 | 出かける | 102 | 冷たい | 21 |
|---|---|---|---|---|---|---|
| 83 | 別れ | 102 | 驚く | 99 | 口惜しい | 20 |
| 84 | 貴女 | 99 | 願う | 97 | 荒々しい | 19 |
| 85 | 役人 | 99 | 贈る | 95 | 細かい | 19 |
| 86 | 命婦 | 96 | 上がる | 94 | 思わしい | 19 |
| 87 | 遊び | 94 | 信じる | 94 | 暑い | 19 |
| 88 | 陛下 | 93 | 進む | 93 | 楽しい | 18 |
| 89 | 身体 | 92 | 劣る | 90 | 詳しい | 18 |
| 90 | 部屋 | 91 | 喜ぶ | 89 | 物足りない | 18 |
| 91 | 容貌 | 89 | 起こす | 89 | 心安い | 16 |
| 92 | 心持ち | 87 | 失う | 89 | 気味悪い | 15 |
| 93 | 後宮 | 86 | 現われる | 88 | 卑しい | 14 |
| 94 | 親王 | 86 | 折る | 88 | 片腹痛い | 14 |
| 95 | 高官 | 85 | 添う | 88 | 眠い | 14 |
| 96 | 童女 | 85 | 堪える | 87 | 憂い | 14 |
| 97 | 愛情 | 84 | 比べる | 86 | 愛らしい | 13 |
| 98 | 田舎 | 84 | 近づく | 85 | 貧しい | 12 |
| 99 | 美貌 | 84 | 使う | 85 | 気安い | 11 |
| 100 | 兄弟 | 81 | 通う | 85 | 愚かしい | 11 |